\documentclass[aps,prb,twocolumn,superscriptaddress]{revtex4}
\usepackage{graphicx}

\begin{document}


\title{Nanoscale Suppression of Magnetization at Atomically Assembled Manganite Interfaces}


\author{J. J. Kavich}
\affiliation{Department of Physics, University of Illinois at Chicago, Chicago, IL 60607}
\affiliation{Advanced Photon Source, Argonne National Laboratory, Argonne, IL 60439}


\author{M. P. Warusawithana}
\affiliation{Department of Physics, University of Illinois at Urbana-Champaign, Urbana, IL 61801}
\author{J. W. Freeland}
\affiliation{Advanced Photon Source, Argonne National Laboratory, Argonne, IL 60439}
\author{P. Ryan}
\affiliation{Advanced Photon Source, Argonne National Laboratory, Argonne, IL 60439}
\author{X. Zhai}
\affiliation{Department of Physics, University of Illinois at Urbana-Champaign, Urbana, IL 61801}
\author{R. H. Kodama}
\affiliation{Department of Physics, University of Illinois at Chicago, Chicago, IL 60607}
\author{J. N. Eckstein}
\affiliation{Department of Physics, University of Illinois at Urbana-Champaign, Urbana, IL 61801}

\date{1 April 2007}

\begin{abstract}
Using polarized X-rays, we compare the electronic and magnetic properties of a La$_{2/3}$Sr$_{1/3}$MnO$_{3}$(LSMO)/SrTiO$_{3}$(STO) and a modified LSMO/LaMnO$_{3}$(LMO)/STO interface. Using the technique of X-ray resonant magnetic scattering (XRMS), we can probe the interfaces of complicated layered structures and quantitatively model depth-dependent magnetic profiles as a function of distance from the interface. Comparisons of the average electronic and magnetic properties at the interface are made independently using X-ray absorption spectroscopy (XAS) and X-ray magnetic circular dichroism (XMCD). The XAS and the XMCD demonstrate that the electronic and magnetic structure of the LMO layer at the modified interface is qualitatively equivalent to the underlying LSMO film. From the temperature dependence of the XMCD, it is found that the near surface magnetization for both interfaces falls off faster than the bulk. For all temperatures in the range of 50K - 300K, the magnetic profiles for both systems always show a ferromagnetic component at the interface with a significantly suppressed magnetization that evolves to the bulk value over a length scale of $\sim$1.6 - 2.4 nm. The LSMO/LMO/STO interface shows a larger ferromagnetic (FM) moment than the LSMO/STO interface, however the difference is only substantial at low temperature.
\end{abstract}


\maketitle


\section{Introduction}
There has been extensive studies of the bulk properties of manganites; however, recently more attention has been focused on probing the properties of interfaces, which are important structures for understanding spin-polarized transport in strongly correlated materials. In manganites, for example, interfaces and free surfaces are known to behave much differently than bulk samples of the same material, and we can expect that many factors can have a profound impact on the balance of competing interactions governing the ground state electronic and magnetic structure. In simple metallic systems, deviations from bulk behavior are quickly screened and highly localized at the surface. However, due to intrinsic disorder and strong interactions, the cross-over from surface to bulk behavior in complex oxides can be expected to be inhomogeneous and extend spatially over longer length scales. This cross-over region directly impacts our understanding of spin-dependent transport in strongly correlated oxide systems.

Bulk, hole-doped La$_{1-x}$Sr$_{x}$MnO$_{3}$ (LSMO) with x$\sim$1/3, for example, is ferromagnetic (FM) and nearly half-metallic at low temperatures. Experiments indicate that the spin-polarization of these materials is at least 0.82\cite{ODonnellAPL} with some evidence for it being completely spin-polarized \cite{ParkNature, BowenAPL, NadgornyPRB} (compared with a value between $\sim$0.45 and $\sim$0.35 for FM Fe and Co)\cite{SoulenScience, ViretMMM}. Using this information, several groups have investigated magnetic tunnel junctions (MTJs) using manganite films as electrodes and have observed a significant reduction in tunneling magneto-resistance (TMR) between 200K and 270K \cite{ODonnellAPL, ViretMMM, BowenAPL, ObataAPL, KwonAPL}. Surface sensitive spin-resolved photoemission studies by Park, et al., indicated that a strong possibility for this loss of TMR is a degraded average magnetization in the surface region\cite{ParkPRL}. Experiments of this kind are surface sensitive, though, and cannot give a quantitative magnetic profile due a probing depth of $\sim$5 $\AA$. 

Attempts have also been made to influence the magnetic properties of an LSMO/SrTiO$_{3}$(STO) interface \cite{IshiiAPL, YamadaScience}. Using pulsed laser deposition (PLD), two unit cells (u.c.) of the undoped parent compound LaMnO$_{3}$ (LMO) were inserted to create an LSMO/LMO/STO structure. Magnetic second-harmonic generation performed on those structures indicated that the magnetization is enhanced at the interface. MTJs made from the two interfaces demonstrated improved TMR from 50$\%$ to 170$\%$\cite{IshiiAPL,YamadaScience}. The TMR of these structures, however, were significantly lower than other published values of unmodified tunnel junctions using STO as a tunnel barrier\cite{ODonnellAPL, ViretMMM, BowenAPL, ObataAPL, GarciaPRB}. Answering, then, the questions of exactly how the magnetic properties of the system change from surface to bulk behavior and whether or not localized doping can dramatically affect the magnetization can enhance our understanding of the properties of manganites and perhaps be applicable in general to strongly correlated electron materials.

In this article, we present a direct measurement of the depth-dependent magnetization profile as a function of temperature in LSMO/STO and modified LSMO/LMO/STO structures. Combined with X-ray absorption spectroscopy (XAS) and X-ray magnetic circular dichroism (XMCD) we can also compare the local electronic and magnetic structure of each system. The electronic and magnetic structure of the LMO layer introduced at the interface appears to be qualitatively equivalent to the underlying LSMO film. Our results show a larger net magnetization in the LSMO/LMO/STO interface, however only at low temperature. The room temperature magnetic profiles are nearly identical for both interfaces, indicating that reducing the density of holes localized at the interface (i.e., inserting 2 u.c. of LMO) does not have a significant effect on the room temperature interface magnetism. At all temperatures both interfaces show a suppressed interface magnetization which evolves continuously to the bulk value over a length scale of $\sim$1.6 - 2.4 nm (4-6 u.c.).

\section{Experiment}
\subsection{Samples}
\begin{figure}[h]
\begin{tabular}{lr}
\includegraphics[scale=.45]{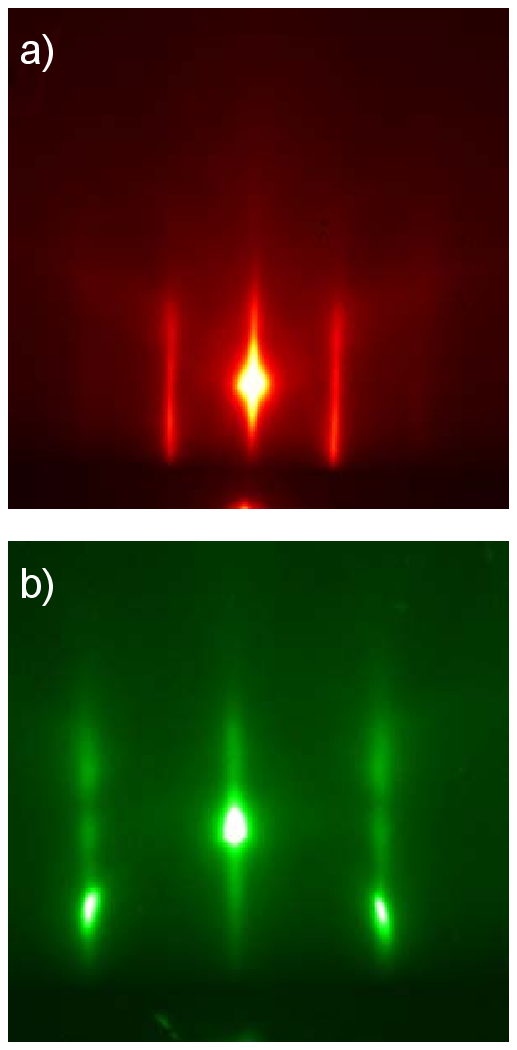} & \includegraphics[scale=.3]{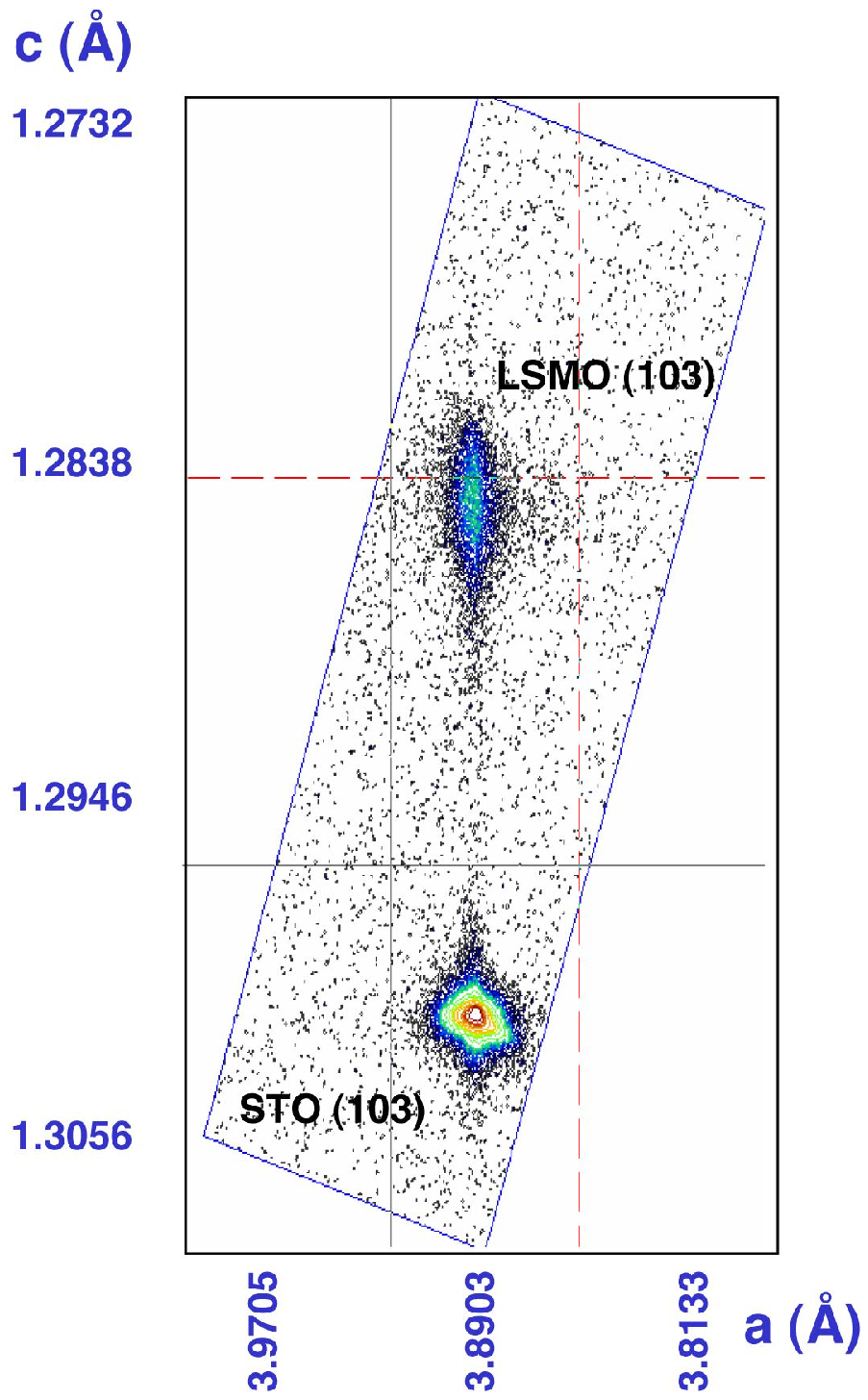} \\
\end{tabular}
 \caption{(Color Online) Left: RHEED images demonstrate that the LSMO growth mode in the ALL-MBE process is indeed 2-dimensional. Pattern (a) was taken at the topmost LSMO layer near the end of the growth process while (b) shows the initial RHEED pattern at the onset of growth nearest the STO substrate. Right: Reciprocal space map illustrating that the in-plane lattice constants on identically grown films of 150 u.c. thickness are completely strained to the STO substrate. The data is taken across the (103) which explains the factor of three difference in the c-axis direction.}
\protect\label{RHEEDnRSP}
\end{figure}

The samples used in our experiment were grown using ozone-assisted atomic layer-by-layer molecular beam epitaxy (ALL-MBE) \cite{ODonnellAPL, EcksteinAnn} on a TiO$_{2}$ terminated STO (100) substrate. The atomic beam fluxes were calibrated to better than 1$\%$ accuracy using Rutherford back scattering spectroscopy and X-ray film thickness oscillations on separate films. The growth was carried out at a substrate temperature of 680$^{\circ}$C and an ozone pressure of 2x10$^{-6}$ torr using flux matched co-deposition. Throughout the LSMO growth, in-situ reflection high energy electron diffraction (RHEED) measurements (shown in Fig.\ \ref{RHEEDnRSP}) indicated intensity oscillations of the specular reflection characteristic of a 2-dimensional growth mode. Immediately after the growth of a 75 u.c. ($\sim$29.7 nm) LSMO layer, one film was capped with 2 u.c. ($\sim$0.8 nm) of STO (LSMO/STO interface).  The other sample was grown under the same conditions with 2 u.c. ($\sim$0.8 nm) of LMO and 2 u.c. of STO on top of a 150 u.c.($\sim$59.4 nm) LSMO layer (LSMO/LMO/STO interface). Despite the difference in film thickness, both samples demonstrate a bulk-like magnetic Curie temperature, T$_{c}$ $\sim$ 360 K.

The electronic and magnetic properties of oxide systems are strongly coupled to lattice distortions, and this makes strain states in thin films a significant concern. Values quoted in the literature for relaxation effects in PLD grown films are varied. One group reports relaxation effects beginning at thicknesses around 100 u.c.\cite{ArutaPRB} and other work shows fully strained in-plane LSMO lattices with thicknesses up to 150 u.c.\cite{GosnetJAP}, both on STO substrates. Strain, though, is a property highly dependent on growth mode. X-ray diffraction on identically grown ALL-MBE LSMO films of 150 u.c. show nearly identical in-plane lattice constants indicating that the LSMO grows pseudomorphic to the STO substrate and is completely strained in our samples. The reciprocal space plots of the diffraction data is presented in Fig.\ \ref{RHEEDnRSP}. Due to the overall in-plane tensile strain of the LSMO, variations in the c-axis lattice constant must be noted. The assymetric peak in the reciprocal space plot shows that there is a contraction in the c-axis lattice constant.  

In previous studies of single-crystal MTJs comprised of a thin epitaxial CaTiO$_{3}$ or SrTiO$_{3}$ tunnel barrier sandwiched between two La$_{0.67}$Sr$_{0.33}$MnO$_{3}$ electrodes, large values of low temperature TMR exceeding 400$\%$ were obtained \cite{ODonnellAPL}. Using the Juliere model, this TMR corresponds to a spin-polarization of $\sim$82$\%$. The structures used to measure the magnetization profiles in this study were prepared identically to those MTJs.

\subsection{X-ray Measurement}
To characterize the interface magnetism, two complementary X-ray techniques were used at beamline 4-ID-C of the Advanced Photon Source: X-ray magnetic circular dichroism (XMCD) \cite{ChenPRL} and X-ray resonant magnetic scattering (XRMS) \cite{KaoPRL}. Both XMCD and XRMS measurements were taken simultaneously across the Mn L$_{2,3}$ edges at a fixed incident angle of 11$^{\circ}$ while varying the temperature over a range of 50 - 300 K using in-plane fields of 500 Oe to saturate the magnetic moment of the sample.

\begin{figure}
 \includegraphics[scale=0.65]{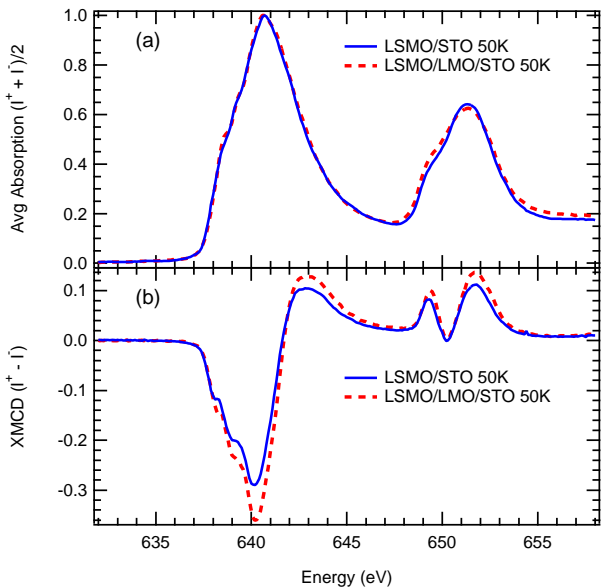}
 \caption{(Color Online) Top(a): Average absorption across the Mn L$_{3,2}$ edges showing identical electronic structure for each interface.  Bottom(b): The LSMO/LMO/STO interface shows an increase in the near-surface magnetization at low temperature. The identical line shape in the XMCD spectra indicates that the ratio of the spin to orbital moment, ($\mu$$_{s}$/$\mu$$_{l}$), is the same for each interface.}
\protect\label{absXMCD}
 \end{figure}

In general, transition metal elements have core level excitation energies corresponding to wavelengths in the soft X-ray region. The radiation interacts with the solid by exciting core level electrons into the vacuum according to dipole selection rules. Outer-shell valence electrons then continuously recombine with the core hole through various channels attempting to minimize the energy of this excited state. The resulting photocurrent is measured through total electron yield (TEY) by replacing the valence state vacancies. The well-established XMCD technique, measured through TEY, probes spin-dependent absorption because the angular momentum vector from the different circularly polarized X-rays interacts preferentially with electrons of opposite spin states. The photocurrents due to right and left circular polarization are measured independently and are denoted by I$^{+}$ and I$^{-}$, respectively. The X-ray absorption (XAS) and XMCD are then calculated by averaging and taking the difference of the photocurrent signals from each polarization, (I$^{+}$ + I$^{-}$)/2 and (I$^{+}$ - I$^{-}$), respectively. It is important to note that the XAS and XMCD probes all the available unoccupied states, including both the localized (t$_{2g}$) and de-localized (e$_{g}$) electrons. Transport measurements, however, only yield information pertaining to the de-localized mobile electron states. Also, the XMCD in the TEY is proportional to an average near-surface magnetic moment because the measured signal is weighted by an exponential with a decay length (mean free escape depth of electrons emitted from the surface) of a few nm for LSMO. The low temperature XAS and XMCD are presented in Fig.\ \ref{absXMCD} and will be examined in detail in section\ \ref{Results}.

To map the magnetization profile of the interface, we utilize XRMS which probes the spatial dependence of the magnetization along the axis normal to the film surface. If we consider an atomic model, the origin of the spin-dependent scattering (or polarization dependent index of refraction) is due to a spin-orbit correction in the Hamiltonian\cite{ArgyresPR}. This term effectively adds a charge-magnetic interference term in the scattering amplitude. At the resonant energy of a FM atom, however, this effect is enhanced and the magnetic correction term becomes comparable to the pure charge scattering\cite{KaoPRL}. The resulting XRMS spectra (I$^{+}$ - I$^{-}$) can be understood as the charge-magnetic interference term in the scattering amplitude where the pure charge scattering has been subtracted off. In the soft X-ray regime, the longer wavelengths (compared to the unit cell lattice dimensions) allow us to calculate the specular scattered field intensities using a magneto-optical boundary matrix formalism\cite{ZakMMM}. Then, the charge-magnetic term in the scattering amplitude can be interpreted as interference between specularly reflected X-rays from the chemical boundaries and those reflected from magnetic planes. The boundary matrix formalism takes into account multiple scattering events while allowing a straight-forward construction of complicated, idealized, layered structures.

As a note, depth-dependent information can be obtained from XRMS in two ways. In standard X-ray diffraction, the depth-dependent charge density can be obtained from $\theta$/2$\theta$ angle dependent scans. Depth-dependent information from fitting the observed finite thickness oscillations relies on changing the q-vector (the momentum transfer). It is important to note, though, that scans of this type are taken at an off-resonant fixed energy, meaning that the index of refraction for materials are constant and usually differ from unity by a few parts in a thousand in the X-ray regime. However, q$_{z}$ is also dependent on incident photon energy through q$_{z}$ = 4$\pi$sin($\theta$) / $\lambda$. Near a resonance, however, the index of refraction of a material is very strongly energy dependent, and so will the interference conditions for specular scattering. We rely on this strong energy dependence across a resonant condition for the necessary contrast to obtain spatial information.

\section{XRMS Modeling}
\begin{figure}
\includegraphics[scale=.65]{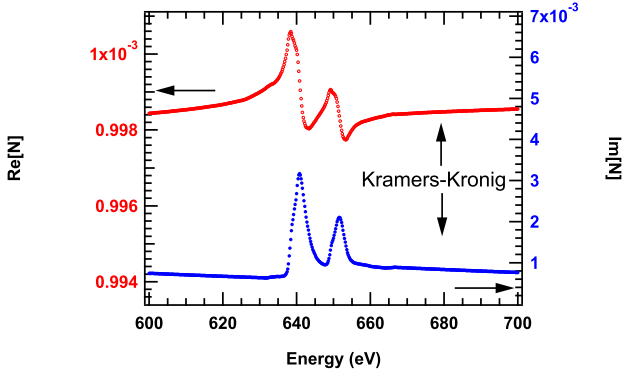}
 \caption{(Color Online) The bottom curve is the average low temperature absorption data illustrated in Fig.\ \ref{absXMCD} spliced into an off-resonant calculation of Im[N(E)]. The top curve is Re[N(E)] which is constructed using a numerical Kramers-Kronig transformation. The entire function N(E) is then described as Re[N(E)] + iIm[N(E)]. The magneto-optical coefficient Q(E) is constructed in the same way using the low temperature XMCD which is proportional to Im[Q(E)].} 
\protect\label{Nplot}
\end{figure}

Due to the long wavelength of the soft X-rays (the range from 500 eV - 800 eV covering both the Oxygen K-edge and the Manganese L-edge corresponds to $\sim$ 2.48 nm - 1.55 nm wavelengths) with respect to the atomic spacings (in perovskite manganites $\sim$0.4nm). The sample can then be described as a continuous medium and the scattering intensity can be calculated using the same formalism as the magneto-optical Kerr effect\cite{ZakMMM}. The starting point to understanding XRMS rests of the resonant behavior of the dielectric tensor. We have already stated that the XRMS results from a polarization dependent interaction between a photon and the magnetic moment of a magnetic material.  Mathematically, this is equivalent to using one polarization of light and reversing the direction of the magnetic moment of the sample. When the magnetic moment of the film lies in the plane, the dielectric tensor contains the following elements:

\begin{equation}
\epsilon (E)=N(E)^2\left( {\matrix{{1}&{0}&{iQ(E)}\cr
{0}&{1}&{0}\cr
{-iQ(E)}&{0}&{1}\cr
}} \right),
\vspace{0.1in}
\end{equation}
where N(E) is the energy dependent index of refraction, and Q(E) is the magneto-optical coefficient. The direction of the moment is identified by the position of Q(E) in the tensor and the strength of scattering is proportional to the magnitude of Q(E). To determine the proper dielectric tensor requires measurement of the N(E) and Q(E). Using the fact N(E) and Q(E) are complex functions and that the imaginary parts correspond to the absorption and XMCD, respectively, we can reconstruct the entire function using a numerical Kramers-Kronig transformation. This is done quantitatively by scaling the low temperature absorption and XMCD data and then splicing it into the tabulated data for the non-resonant scattering factors illustrated in Fig.\ \ref{Nplot}. At low temperatures, well below T$_{c}$, we measure the largest XMCD signal and we associate that with the bulk magnetization and the largest value of Q(E). For simplicity, we have assumed that the material is electronically homogeneous and isotropic (i.e. the diagonal elements of the dielectric tensor are equal) which is not always true in the case of complex oxides.

To construct the scattering from the corresponding dielectric tensor, we need to determine the corresponding reflectivity of the sample at a given angle $\theta$ and energy E represented as:

\begin{equation}
R(\theta,E)=\left( {\matrix{{r_{ss}(\theta,E)}&{r_{sp}(\theta,E)}\cr
{r_{ps}(\theta,E)}&{r_{pp}(\theta,E)}\cr
}} \right)
\end{equation}
where $r_{ss}$ and $r_{pp}$ are the reflection coefficients for s and p polarized light (perpendicular and parallel to the plane of incidence) and $r_{sp}$ and $r_{ps}$ are magnetic reflectivity terms that mix the s and p polarization states. The incoming circularly polarized photon is described as
\begin{equation}
E_{In}^{\pm}=A\left( {\matrix{1\cr
{\mp i}\cr
}} \right),
\end{equation}
the helicity dependent scattered intensity is then
\begin{equation}
I^{\pm} = \left| {R(\theta,E) \cdot E_{In}^{\pm}} \right|^2
\end{equation}
From this, the sum and difference spectra
are then determined as
\begin{equation}
\left(I^++I^-\right)=A^{2}\left[\left| {r_{ss}} \right|^2+ \left|
{r_{pp}} \right|^2+\ldots\right] and
\protect\label{chemscatt}
\end{equation}
\begin{equation}
\left(I^+-I^-\right)=-4A^{2}
Im\left[r_{ss}^{*}r_{sp}+r_{pp}r_{ps}^*\right],
\protect\label{magscatt}
\end{equation}
where higher order magnetic terms in the sum are small and can be
ignored. Equations\ \ref{chemscatt} and
\ref{magscatt} then show clearly that ($I^++I^-$) is purely chemical while
($I^+-I^-$) contains both chemical ($r_{ss}$ and $r_{pp}$) and magnetic
scattering
contributions ($r_{sp}$ and $r_{ps}$), which directly indicates that it is
not purely magnetic in origin.

\begin{figure}
\includegraphics[scale=.6]{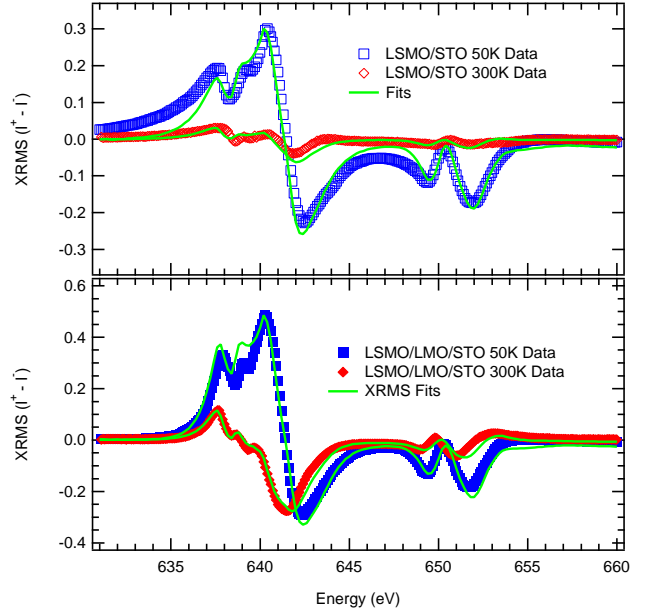}
 \caption{(Color Online) Data and characteristic fits for low and high temperatures for each of the interfaces studied. Taking the difference in circularly polarized scattering is equivalent to subtracting the pure charge scattering. The remaining charge-magnetic interference term is modeled. The line shape of the XRMS depends solely on the depth-dependent magnetic profile.} 
\protect\label{XRMSfits}
\end{figure}

For the case of a magnetic surface, the analytical form for the elements composing R($\theta$,E) are known.  However, due to the strong multiple scattering in the soft X-ray regime, analytical forms are not simple to construct even for a single layer film\cite{YangJAP}.  In order to simulate the XRMS data, we follow the MOKE formalism using boundary matrices as outlined by Zak, Moog, and Bader\cite{ZakMMM}. By representing the sample in a slab structure and using the equations above, the scattered intensity can be calculated and compared with the experimental data.

\begin{figure}
\includegraphics[scale=.6]{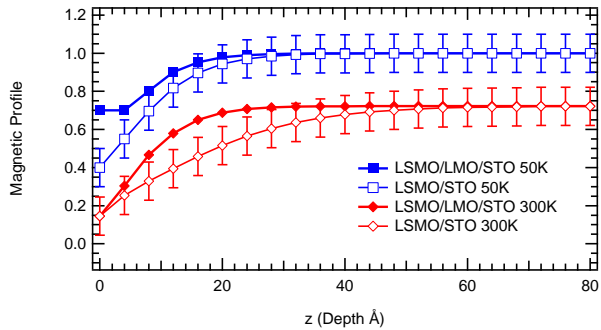}
 \caption{(Color Online) After modeling many different functional dependencies, the depth-dependent magnetic profiles are presented for both the LSMO/LMO/STO and the LSMO/STO interfaces. z=0 describes the LSMO/STO and LMO/STO interface. The modified LMO/STO interface shows increased magnetization at low temperature; however, the surface magnetization M$_{sur}$ = M$\mid$$_{z=0}$ is still significantly suppressed. } 
\protect\label{MagProfiles}
\end{figure}

To construct the boundary matrices, we use the facts that the quasi-cubic perovskite unit cell contains only one Mn ion, and the distance between possible magnetic scattering planes is fixed by the lattice constant ($\sim$0.4 nm). Explanation of the XRMS depends on fitting the spectra by varying the magnitude Q(E) corresponding to the average magnetization in each magnetic MnO$_{2}$ plane, which is specified by its depth in the sample (i.e., either the LSMO/STO or the LMO/STO capping interface being designated z=0). If the incident angle and chemical boundaries are known, the detailed shape of the XRMS depends only on the shape of the magnetization profile.  The magnetization profile was parameterized by the width of the transition region and magnetization at the interface.  We tested this for a variety of functional forms, but in all cases we required that the profile be continuous and vary in a physical way. This methodology has had remarkable success in fitting the observed data (Fig.\ \ref{XRMSfits} and leads directly to the depth-dependent magnetic profile (Fig.\ \ref{MagProfiles}). 

\section{Results}\protect\label{Results}
Analysis of the average absorption (XAS) for all temperatures taken across the Mn L$_{2,3}$ edges for the the two different interfaces have nearly identical line shapes (the low temperature data is illustrated in Fig.\ \ref{absXMCD}). The average escape depth of the electrons from the surface is typically no more than 5 nm\cite{GotaPRB}, therefore, the LMO interface layer is at least responsible for 20$\%$ of the total signal. The identical line shape then, indicates that the electronic structure of the unoccupied states in the LMO at the interface is similar to the underlying LSMO. In contrast, the XMCD shows that the magnitude of the near-surface magnetization for the LSMO/LMO/STO is nearly 10$\%$ larger than the LSMO/STO interface at 50K (Fig.\ \ref{absXMCD}). The XMCD line shapes are otherwise identical and can be scaled to overlay exactly, which demonstrates that the ratio of spin to orbital moments ($\mu$$_{s}$/$\mu$$_{l}$) is the same for both interfaces. From this, we conclude that the local moments are the same. The total average magnetization in the near surface region is proportional to the area bounded by the XMCD spectra\cite{ChenPRL}. Due to the identical line shape, the analysis is simplified and the temperature dependence of the magnetization can be described using the maximum peak height.

\begin{figure}[h]
\includegraphics[scale=.6]{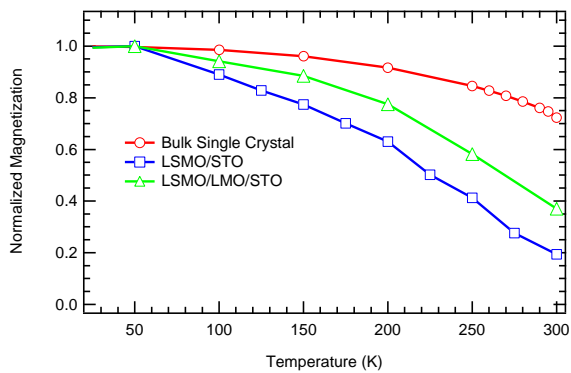}
\caption{(Color Online) The temperature dependence of the XMCD peak height (proportional to the near-surface magnetization) for the LSMO/STO and the LSMO/LMO/STO interfaces is plotted against the temperature dependence of a bulk single crystal LSMO of the same composition.   The LSMO/LMO/STO interface has a slightly better thermal stability than the LSMO/STO interface; however, it is still roughly 40$\%$ of the bulk value at room temperature.} 
\protect\label{MvsT}
\end{figure}

In Fig.\ \ref{MvsT} we plot the XMCD peak height as a function of temperature for each sample and compare it to the temperature dependence of a bulk LSMO single crystal. The bulk data was measured using SQUID magnetometry\cite{SalamonPCom} on a bulk LSMO single crystal of similar composition (x=0.3), T$_{c}$, and transport characteristics. The XMCD peak heights were scaled to the low temperature bulk value.  From the plot we see that the surface magnetization is falling off faster than the bulk for both interfaces (See Fig.\ \ref{MvsT}). The deviations of the average magnetization for the different interfaces from the bulk show a similar functional dependence on temperature, however, the LSMO/LMO/STO interface decreases more more gradually than the LSMO/STO interface.  This demonstrates that the LSMO/LMO/STO interface has an increased average moment in the near surface region.  This result is very similar, to previous work \cite{ParkPRL}.

The LMO layer in the atomically assembled structure has two significant effects on the system. In the LSMO/LMO/STO sample, the surface magnetization (M$_{sur}$ = M(z=0) defined to be the STO interface) is roughly 70$\%$ of the bulk value at low temperature. At the same time, the low temperature LSMO/STO interface is only 40$\%$ of the bulk value which is consistent with the larger XMCD signal presented in Fig.\ \ref{absXMCD}.  Secondly, the LMO at the interface sharpens the transition region as seen by comparing the slopes in FIG.\ \ref{MagProfiles}. Each profile demonstrates a continuous decrease from bulk ferromagnetism (M=1) to some non-zero value at the interface with a transition region of $\sim$1.5 nm ($\sim$4 u.c.) for the LSMO/LMO/STO interface and $\sim$2.5 nm ($\sim$6 u.c.) for the unmodified interface.  In each case though, the length scale set by the transition region is nearly an order of magnitude larger than the Thomas-Fermi screening length of $\sim$0.3 nm predicted from the density of carriers in x = 0.3 LSMO \cite{HongAPL}.  

The XRMS indicates that both the LSMO/STO and LSMO/LMO/STO interfaces have similar magnetic structure.  There is a measurable increase in the surface moment outside experimental error at low temperature, which is also observed in the XMCD (See Fig.\ \ref{absXMCD}). At high temperature, however, the profiles do not differ more than the experimental error, and each interface converges to approximately the same value of M$_{sur}$ =  M$\mid$$_{z=0}$.  The profiles also demonstrate a reversible nature as a function of temperature. At this point, it is important to clarify the interpretation of the profile and its relationship with the temperature dependence of the peak height in the XMCD. The magnetic profile derived from XRMS is the complete picture, and the value we defined as M$_{sur}$, is the value which can be associated with the surface magnetization acquired from spin-resolved photoemission measurements.  The peak height in the XMCD corresponds, again, to the average magnetization in the near surface region. The complete magnetic profiles at each temperature can be weighted by the electron escape probability and then averaged to reproduce the results from the temperature dependence of the XMCD peak heights, showing consistency between two independent measurements.

\section{Discussion}
Our results are in partial agreement with the work of Ishii\cite{IshiiAPL} and Yamada et al.\cite{YamadaScience}. Their results clearly show an improvement in the modified interface at low temperature. They have reported an improvement in the TMR from 50$\%$ to 170$\%$ but only at low temperatures. The spin-polarization derived from their tunneling measurements converges to zero around the same high temperature value. They still, however, report an enhanced interface magnetization at room temperature which is not supported by our measurements.

Previous studies of magnetic tunnel junctions grown identically to the isolated interfaces presented here have achieved TMR ratios in excess 400$\%$\cite{ODonnellAPL}, even in MTJs without the LMO modification. Other groups have also reported similar values of TMR\cite{ViretMMM, BowenAPL, ObataAPL, GarciaPRB}. We demonstrate that there is an intrinsic loss of FM at the surfaces and interfaces of manganite materials. We show that the LMO enhances the magnetization at low temperature while the interface magnetization converges to nearly the same value at room temperature. This is also consistent with the low temperature magnetization results and the temperature dependence of the spin polarization presented by Ishii, et al.\cite{IshiiAPL}.

For a free LSMO surface the reduced magnetization can be attributed to a number of causes. One of the most likely causes is a preferential filling of the in-plane versus out of plane orbitals in the near surface region.  In the bulk there is an equal probability that an electron will occupy either the e$_{g}$$_{(3z^{2} - r^{2})}$ or the e$_{g}$$_{(x^{2} - y^{2})}$ orbitals. The broken symmetry at the surface results in a loss of coordination of the orbitals in the direction normal to the surface. This may cause a reduction in bandwidth of the e$_{g}$$_{3z^{2} - r^{2}}$ orbitals giving the e$_{g}$$_{(x^{2} - y^{2})}$ a preferred occupancy. An increase in the in-plane orbital occupancy would justify a reduced inter-planar hopping which will frustrate double exchange in that direction and improve the chances of the superexchange interaction to produce a-type antiferromagnetic order in the layers closest to the surface.  This idea is supported by calculations of magnetic surface states\cite{ZeniaPRB} of manganites. 

This same effect of preferential orbital occupancy could simply be a result of the strain state as well.  In strained thin film samples, even though the in-plane lattice constants are quite controlled, the c-axis lattice constants are free to compenstate.  As seen by the reciprocal space plots in Fig.\ \ref{RHEEDnRSP}, there is a distribution in the direction normal to the surface.  In general, these films are under tensile strain which would cause a small reduction in the c-axis lattice constant.  This would also give a preferential filling to the in-plane orbitals. Intermixing of Sr atoms into the LMO layer is not likely in these samples because the kinetics of atomic diffusion perpendicular to the surface occur at much higher temperatures than were used in the ALL-MBE growth process. This is supported by cross sectional TEM images which do not show any indication of atomic interdiffusion \cite{PalanisamiPRB}. 

Another result which supports our conclusion is recent theoretical work demonstrating a mixed phase of CE type AFM ordering and FM at manganite insulator interfaces \cite{BreyPRB}.  This is also consistent with our measurements since the XRMS profiles integrate over lateral variations.  It is not possible with this measurement alone to determine the in-plane magnetic order at the interface. It is very likely that the interface region is inhomogeneous, but we show that the surface always has some FM component. One possibility that cannot be dismissed is a percolation effect in which there is a mixed phase FM order in an increasing AFM background as the interface is approached.

The magnetic profile reveals that the LMO in the LSMO/LMO/STO is dominantly FM, whereas truly bulk LMO is antiferromagnetic (AFM). The average absorption indicates that the electronic structure of unoccupied states is qualitatively the same as the underlying LSMO, suggesting that doped holes can, indeed, diffuse into the LMO. This result fits nicely with calculations that study the effect of carrier concentration\cite{BreyPRB, ZeniaPRB} on the magnetic ground state. At low-temperatures, there is a high number of mobile carriers in the LSMO because it is in a FM metallic state. Thin film LSMO has been shown to have a metallic-like screening length of $\sim$0.3 nm, corresponding to a single unit cell\cite{HongAPL}. The diffusion length of the carriers is then smaller because they screen themselves.  As the temperature is increased nearer to the FM Curie temperature (the loss of FM is correlated with an increase in resistivity) would mean that the diffusion lengths of doped holes could increase due the reduced screening effect of a fraction of the carriers becoming more localized. 

The interpretation of our results and the comparison to tunneling measurements must be addressed as well. Our results demonstrate that the interface at low temperatures in the unmodified interface is only 40$\%$ of the bulk value, while the magnetic tunnel junctions of the same materials demonstrate a spin-polarization of 80$\%$ and from spin-polarized photoemission experiments it can be deduced that these materials are nearly half-metallic, meaning almost 100$\%$ of the carriers are polarized.  However, this is not necessarily a discrepancy.  Tunneling and photoemission methods reliably measure regions of the surface or interface which are predominately metallic.  Metallic regions, though, may be only a fraction of the interface and not give insight into the whole picture. From this, we argue, that if a portion of the interface were indeed non-ferromagnetic and insulating (which is supported by calculucations already discussed), our measurements are consistent with transport and spin-polarized photoemission.  We show that the ferromagnetic metallic component is roughly 40$\%$ of the entire interface. Measurements probing the metallic portion of the interface would then measure a very high degree of spin-polarization.  

\section{Conclusion}
The magnetic profiles and X-ray spectroscopy both support the idea that the LMO layer is chemically and magnetically similar to the underlying LSMO, suggesting that the doped hole carriers can diffuse over length scales of at least 2 u.c.  This is supported by the width of the transition region of the magnetic profiles across each interface ($\sim$3 u.c. - 6 u.c).  The magnetic behavior of the interface, when carefully probed, demonstrates that the smoothly varying surface magnetization is still significantly degraded at all temperatures.  Extensive work on the effect of variation of the tunnel barrier in these systems also shows that, in all cases, the temperature dependence of the TMR is similar \cite{GarciaPRB}, and from that it can be deduced that the normalized magnetization profiles for various LSMO / tunnel barrier interfaces will be in qualitative agreement with our results.

\begin{acknowledgments}
The use of the Advanced Photon Source was supported
by the U.S. Department of Energy, Office of Science, Of-
fice of Basic Energy Sciences, under contract W-31-109-
ENG-38.
\end{acknowledgments}

\bibliography{LSMOalloy_PRB}

\end{document}